\documentclass[aps,reprint,amsmath,amssymb,floatfix,longbibliography,groupedaddress]{revtex4-2}

\usepackage{graphicx}
\usepackage{amssymb}
\usepackage{amsmath}
\usepackage{bm}
\usepackage{xcolor}
\usepackage{verbatim}
\usepackage{dcolumn}
\usepackage{multirow}
\usepackage{url}
\usepackage{hyperref}

\begin{document}

\title{Formation of bosonic $^{23}$Na$^{41}$K Feshbach molecules}

\author{Sungjun Lee}
\thanks{These authors contributed equally to this work.} 
\author{Younghoon Lim$^{*}$}
\altaffiliation[Current address: ]{3rd R\&D Institute, Agency for Defense Development,
Daejeon 34186, Republic of Korea.} 
\author{Jongyeol Kim}
\author{Jaeryeong Chang}
\author{Jee Woo Park}
\email{jeewoopark@postech.ac.kr}
\affiliation{Department of Physics, Pohang University of Science and Technology,
Pohang 37673, Republic of Korea}

\begin{abstract}
Ultracold Feshbach molecules are a crucial intermediate step for the creation of quantum degenerate gases of strongly dipolar molecules. After coherent transfer to the rovibrational ground state, these dimers can realize stable dipolar gases with strong, tunable long-range interactions. Here, we report the creation of bosonic $^{23}$Na$^{41}$K Feshbach molecules by radio-frequency (RF) association. An RF pulse applied on the molecular side of an interspecies Feshbach resonance at 73.6(1)~G associates up to $1.1(1)\times10^4$ molecules from a thermal mixture of $^{23}$Na and $^{41}$K atoms. Measurements of the binding energy reveal a broad resonance width of 5.1(2)~G, facilitating robust control over interspecies interactions. The molecule lifetime in the presence of background atoms exceeds 2~ms and increases to approximately 7~ms after removal of $^{23}$Na. These results constitute a key step toward the production of ultracold $^{23}$Na$^{41}$K ground state molecules and for pursuing shielded evaporative cooling in the dipolar regime.
\end{abstract}

\maketitle

Ultracold dipolar molecules have emerged as a powerful platform for quantum science and technology\cite{Carr2009Review, Bohn2017Review, Langen2024, Cornish2024}. Their large electric dipole moments generate strong, tunable, and anisotropic long-range interactions, granting access to many-body phenomena beyond the reach of contact interactions, including dipolar crystals~\cite{Buchler2007molecules}, quantum spin liquids~\cite{Yao2018Spin}, and the quantum simulation of extended Hubbard and lattice spin models~\cite{CapogrossoSansone2010molecules, Micheli2006molecules}. Leveraging their rich internal structure of vibrational, rotational, and hyperfine states, dipolar molecules support dipole-mediated quantum information protocols~\cite{DeMille2002quantum, Ni2018logicgate, Picard2025iSWAP} and provide exquisite sensitivity for precision tests of fundamental symmetries and physics beyond the Standard Model~\cite{DeMille2024Review}.

The creation of stable, quantum degenerate gases of ground state molecules is the starting point for many of these applications. The established route proceeds by associating weakly bound bialkali dimers near an interspecies Feshbach resonance, followed by coherent transfer to the rovibrational ground state via stimulated Raman adiabatic passage (STIRAP). Since the first creation of fermionic $^{40}$K$^{87}$Rb molecules~\cite{Zirbel2008KRb, Ni2008KRb}, this approach has been successfully applied to an expanding range of bialkali species~\cite{Takekoshi2014RbCs, Molony2014RbCs, Park2015Na40K, Guo2016NaRb, Rvachov2017NaLi, Voges2020Na39Kground, Cairncross2021Ni, Stevenson2023NaCs, He2024LiK, Zamarski2025KCs}. Advances in atomic and molecular control have enabled the creation of a quantum degenerate Fermi gas of $^{40}$K$^{87}$Rb molecules~\cite{DeMarco2019KRb}, while techniques of dipolar shielding and molecular evaporative cooling have led to the realization of a degenerate Fermi gas of $^{23}$Na$^{40}$K~\cite{Schindewolf2022shielding} and molecular Bose-Einstein condensates of $^{23}$Na$^{133}$Cs~\cite{Bigagli2024NaCsBEC} and $^{23}$Na$^{87}$Rb~\cite{Shi2025NaRb}.

Among experimentally realized bialkali molecules, NaK offers several notable advantages. It possesses a substantial electric dipole moment of about 2.7~D~\cite{Gerdes2011}, making it a promising platform for studies of strongly interacting dipolar systems~\cite{Schmidt2022}. NaK molecules in their absolute ground state are stable against exothermic exchange reactions~\cite{Zuchowski2010dimer, Park2015Na40K}, which can suppress two-body losses and potentially enhances the stability of the molecular sample, for instance, in doubly occupied sites of optical lattices or optical tweezers. Furthermore, NaK offers both bosonic and fermionic isotopologues, allowing the exploration of bosonic and fermionic dipolar many-body phenomena as well as comparative studies of how quantum statistics influence ultracold chemical reactions~\cite{Liu2024},\cite{Gersema2021Ospelkaus}. While bosonic $^{23}$Na$^{39}$K Feshbach molecules have been realized~\cite{Voges2020Na39K}, the $^{23}$Na$^{41}$K isotopologue is particularly promising due to favorable inter- and intraspecies scattering lengths of $^{23}$Na and $^{41}$K~\cite{Chang2024NaK}. These properties enable efficient sympathetic cooling and the creation of large, ultracold $^{23}$Na-$^{41}$K mixtures, providing an ideal starting point for the realization of high phase-space density molecular gases.

Here, we report the creation of ultracold gases of weakly bound $^{23}$Na$^{41}$K bosonic Feshbach molecules. A radio-frequency (RF) pulse applied on the molecular side of an interspecies $s$-wave Feshbach resonance at 73.6(1)~G associates over $10^{4}$ dimers from a $^{23}$Na-$^{41}$K mixture at temperatures as low as 170~nK. The corresponding phase-space density of the molecular gas is estimated to be 0.03(1). RF spectroscopy maps the molecular binding energy versus magnetic field, precisely characterizing the Feshbach resonance and revealing a broad width of 5.1(2)~G. Near threshold, we systematically explore the association efficiency as a function of temperature, pulse duration, and atom number ratio. We then employ an extended rate-equation model~\cite{Voges2020Na39K}, modified to include molecular losses from collisions with both $^{23}$Na and $^{41}$K, to interpret the observed association dynamics. Remarkably, this extended model quantitatively reproduces the measured lifetimes without any free parameters, relying exclusively on coefficients derived from the association curves. We observe molecular lifetimes exceeding 2~ms over a range of binding energies, extending to approximately 7~ms after removal of $^{23}$Na. These results demonstrate that high phase-space density samples of $^{23}$Na$^{41}$K Feshbach molecules can be produced and quantitatively characterized, providing an excellent starting point for creating ultracold gases of ground state $^{23}$Na$^{41}$K molecules.

\begin{figure}[t]
 \includegraphics[width=0.48\textwidth]{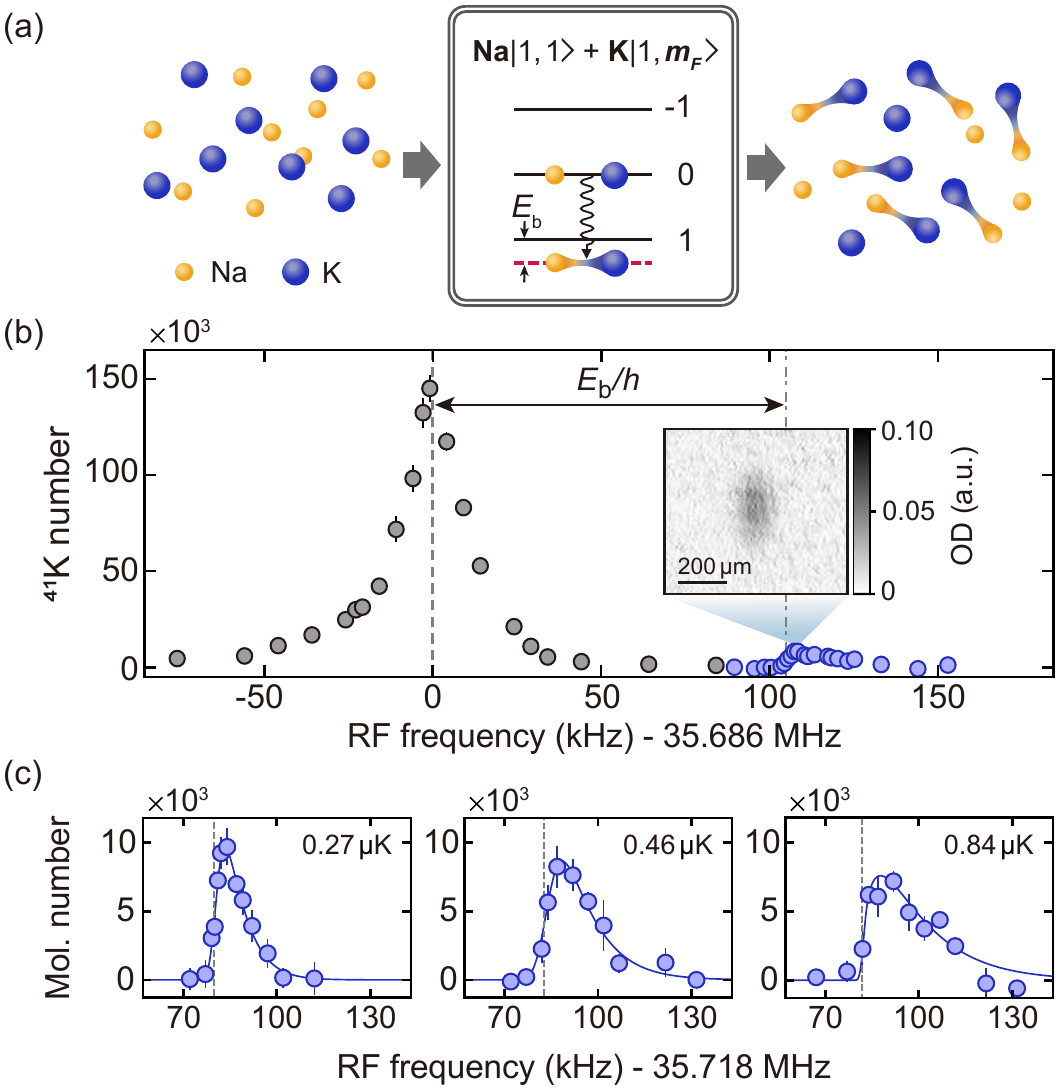}
  \caption{Formation of bosonic $^{23}$Na$^{41}$K Feshbach molecules. (a) Schematic illustration of RF association from free atoms to weakly bound Feshbach molecules. (b) RF association spectroscopy of Feshbach molecules at $B{=}72.3$~G. Gray circles show the number of $^{41}$K $\vert 1,1\rangle$ atoms; the dashed line marks the atomic spin-flip transition at 35.686(1)~MHz. Blue circles show the associated molecule number; the dash-dotted line indicates the fitted molecular resonance at 35.791(1)~MHz. The inset shows an absorption image of molecules taken in trap on the $^{41}$K imaging transition. (c) Number of associated molecules versus RF frequency at $B{=}72.4$~G for mixture temperatures of 0.26(10)~$\mu$K, 0.43(21)~$\mu$K, and 0.73(6)~$\mu$K. Fits to the asymmetric free-to-bound lineshape yield temperatures of 0.27(3)~$\mu$K, 0.46(6)~$\mu$K, and 0.84(12)~$\mu$K, respectively. Error bars denote the standard error of the mean over five independent measurements.}
\label{rf_spectra}
\end{figure}

The experiment begins by preparing ultracold mixtures of $^{23}$Na and $^{41}$K, as described in Refs.~\cite{Chang2024NaK, Chang2025Machine}. For the work presented here, the atoms are confined in an oblate, single-beam 1064~nm optical dipole trap (ODT) in the $\vert F,m_{F} \rangle$ = $^{23}$Na$\vert 1,1 \rangle$ + $^{41}$K$\vert 1,1 \rangle$ hyperfine state combination, where $F$ and $m_F$ denote the total angular momentum and its projection. After evaporative cooling in the ODT, the mixture contains approximately $N_\text{Na}$ = 1--1.5$\times10^{5}$ sodium and  $N_\text{K}$ = 2.5--4$\times10^{5}$ potassium atoms at temperatures between 1.2 and 0.2~$\mu$K.  At the lowest temperature, the trapping frequencies for sodium are $\{\omega_{x}, \omega_{y}, \omega_{z}\}$$\mathrm{_{Na}}$=2$\pi$$\times$$\{18(1) , 8(1), 703(16)\}$ Hz. The atom number ratio and temperature are controlled by varying the number of $^{41}$K atoms captured in the magnetic trap and by adjusting the final ODT depth at the end of evaporative cooling.

To access the interspecies Feshbach resonance previously identified near 73~G for the $^{23}$Na$\vert 1,1 \rangle$ + $^{41}$K$\vert 1,1 \rangle$ collision channel~\cite{Chang2024NaK}, the magnetic field is first ramped to 40~G. Here, the differential Zeeman splitting between the $\vert 1,1\rangle$ $\to$ $\vert1,0 \rangle$ transitions of $^{23}$Na and $^{41}$K is sufficiently large to selectively transfer $^{41}$K atoms to the spectator $\vert 1,0 \rangle$ state via a Landau-Zener (LZ) sweep. Then, the magnetic field is ramped to a final value $B$ on the molecular side of the resonance to perform RF association (see Supplemental Material (SM) for the details of the experimental sequence~\cite{SM}). 

Feshbach molecules are created by using RF photons to coherently couple pairs of colliding atoms to a weakly bound molecular state near a Feshbach resonance. Starting from the $^{23}$Na$\vert 1,1 \rangle$ + $^{41}$K$\vert 1,0 \rangle$ mixture at magnetic field $B$, we apply an RF pulse that drives the $\vert 1,0 \rangle \rightarrow \vert 1,1 \rangle$ transition of $^{41}$K. When the RF frequency is higher than the atomic transition by $E_b/h$, where $E_b$ is the molecular binding energy and $h$ is Planck's constant, the RF field transfers atom pairs into the least-bound molecular state associated with the $^{23}$Na$\vert 1,1 \rangle$ + $^{41}$K$\vert 1,1 \rangle$ threshold (see Fig.~\ref{rf_spectra}(a)).

Figure~\ref{rf_spectra}(b) shows a representative RF spectrum recorded at $B= 72.3$~G. To probe the molecular bound state, we apply a Blackman-shaped RF pulse with a duration of 3.15~ms and Rabi coupling $\Omega=2\pi$~$\times$~9.3~kHz, followed by absorption imaging of $^{41}$K in the $\vert 1,1 \rangle$ hyperfine state. Although weakly bound dimers near threshold can, in principle, be imaged using their constituent atomic transitions, at fields near 73~G neither $^{23}$Na nor $^{41}$K in $\vert 1,1 \rangle$ possesses a closed optical cycling transition. We therefore implement a repumping scheme on $^{41}$K to maintain population in a bright manifold of hyperfine states during imaging, which enhances the observed signal by approximately a factor of two (see details in SM~\cite{SM}). 

Scanning the RF frequency reveals two distinct features corresponding to the atomic spin-flip and molecule association [Fig.~\ref{rf_spectra}(b)]. The molecular peak exhibits a pronounced asymmetry, which arises from the thermal distribution of relative kinetic energies among the colliding atom pairs. To accurately extract the binding energy from the separation between these two features, we fit the molecular spectrum using a free-to-bound transition model that accounts for the phase space overlap of the mixture (see details in SM~\cite{SM}). This yields a binding energy of $E_{b}$/$h$=~105(1)~kHz at $B= 72.3$~G. As this model explicitly accounts for the collision energy distribution, we can also extract the sample temperature from the asymmetric free-to-bound lineshape. Figure~\ref{rf_spectra}(c) shows association spectra at $B= 72.4$~G for varied temperatures. The fitted temperatures agree well with independent time-of-flight measurements, and the binding energies extracted from these fits, $E_{b}$/$h$~=~81(1)~kHz, remain consistent across the temperature range.

\begin{figure}[t]
 \includegraphics[width=0.48\textwidth]{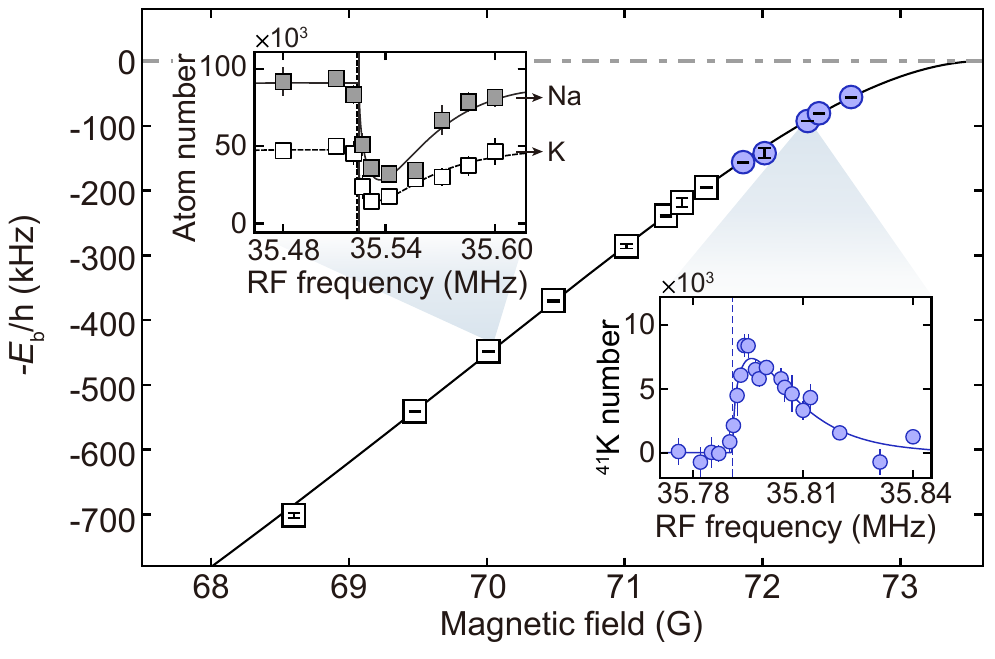}
  \centering
  \caption{Molecular binding energy versus magnetic field. Blue circles and white squares denote measurements from RF association spectroscopy and atom-loss spectroscopy, respectively. The solid line is a fit to the asymptotic binding energy model weighted by the errors. The lower right inset shows the RF association spectrum at $B$ = 72.3~G, yielding $E_{b}$/$h$~=~105(1)~kHz. The upper left inset shows representative atom-loss spectra for Na (gray squares) and K (white squares), giving $E_{b}$/$h$~=~448(1)~kHz. }
\label{binding_energy}  
\end{figure}

Since absorption imaging becomes inefficient for more deeply bound molecules, we determine the binding energy in this regime using atom-loss spectroscopy. To this end, we apply a 200~ms square RF pulse to the atomic mixture, during which Feshbach molecules are formed and subsequently lost from the trap through inelastic collisions with background atoms. We then image the remaining $^{23}$Na and $^{41}$K atoms to obtain a loss spectrum, as shown in the upper inset of Fig.~\ref{binding_energy}. By combining RF association and atom-loss spectroscopy measurements, we map out the molecular binding energy over a broad range of magnetic fields [Fig.~\ref{binding_energy}].

We fit the measured binding energies near threshold with the van der Waals-corrected universal relation $E_b(B){=}\hbar^2/[2\mu (a(B){-}\bar{a})^2]$, where $a(B){=}a_{\mathrm{bg}}[1{-}\Delta/(B{-}B_0)]$~\cite{Rev2010Chin}. Here, $a(B)$ is the interspecies scattering length, $\bar{a}=51.1 a_0$ is the mean scattering length fixed by the van der Waals coefficient of NaK~\cite{Derevianko2001C6}, $a_{\mathrm{bg}}$ is the background scattering length, $B_0$ and $\Delta$ are the resonance position and width, and $\mu$ is the reduced mass of the atom pair. The fit yields $B_0$ = 73.6(1)~G, $\Delta$ = 5.1(2)~G, and $a_{\mathrm{bg}}$ = 235(3)$a_0$, in good agreement with coupled-channel calculations, which predict $B_0 = 73.35$ G, $\Delta = 4.59$ G, and $a_{\mathrm{bg}} = 261.77 a_0$~\cite{Viel2016NaK}. The extracted resonance width is substantially larger than the experimentally measured width via inelastic loss measurement reported in the previous study~\cite{Chang2024NaK}. Importantly, the broad width of this low-field resonance provides a robust and convenient handle for tuning the interspecies interactions over a wide range in our experiments. 

\begin{figure}[t]
\centering
 \includegraphics[width=0.48\textwidth]{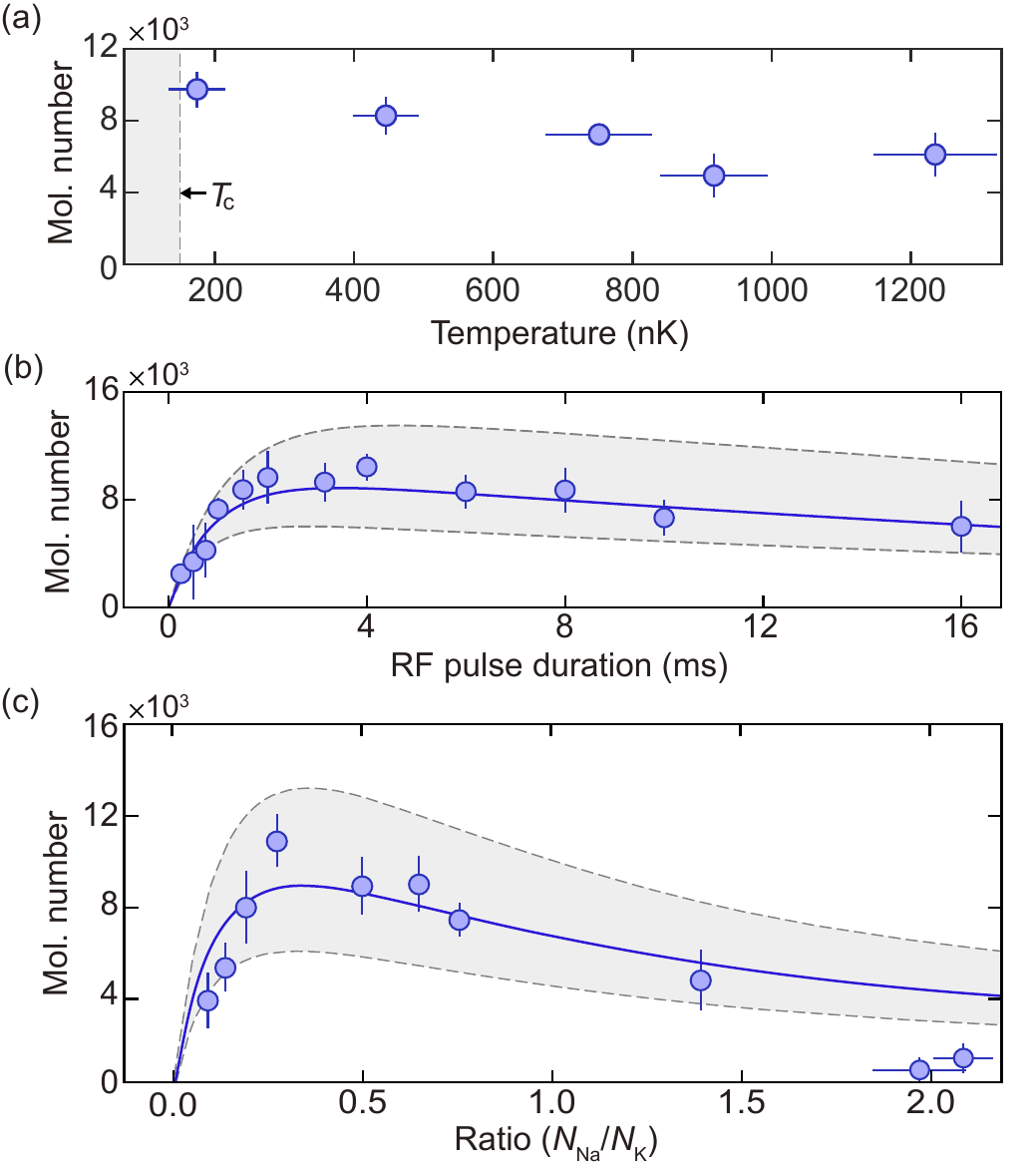}
  \caption{Characterization of molecule association. (a) Number of molecules as a function of temperature. The gray-shaded area indicates the temperature range below the critical temperature of $^{41}$K. The temperatures are extracted from the thermal wings of the $^{41}$K clouds. (b) Molecule number versus RF pulse duration. The maximum number is reached at 3.4~ms, after which the loss becomes dominant over the molecule association. (c) Number of molecules created at various mixture ratios. To ensure an adequate dynamic range for these measurements, the atom numbers are varied over a range of $N_\mathrm{Na}$ = 0.5–3.1$\times10^{5}$ and $N_\mathrm{K}$ = 1.4–5.6$\times10^{5}$ while the typical temperature is kept at $\sim$0.5~$\mu$K. The solid lines represent numerical solutions of the rate equations fitted to the experimental data. The dashed lines are $\pm1\sigma$ confidence intervals The RF pulse power is kept constant for all measurements.}
\label{Optimization}
\end{figure}

To optimize the molecule creation efficiency, we characterize the association process as a function of temperature, RF pulse duration, and atom number ratio. The optimization is performed near the threshold, where the association efficiency depends only weakly on the molecular binding energy. First, we examine the temperature dependence of the molecule number, as shown in Fig.~\ref{Optimization}(a). By adjusting the ODT depth at the end of evaporation, we tune the temperature between 1.2~$\mu$K and 0.2~$\mu$K, while keeping the atom number ratio of 0.4(1). Although the total atom number decreases as evaporation proceeds to lower temperatures, the molecular yield increases due to improved phase space overlap between the two species. The differential gravitational sag between $^{23}$Na and $^{41}$K is about $0.15~\mu$m, which is negligible compared to the sample size along the gravitational direction, $1.3~\mu$m. At the lowest temperatures, we infer a peak molecular phase-space density of 0.03(1). This estimate assumes the molecules are thermalized to the atomic mixture temperature and trapped with frequencies scaled by their polarizability and mass. However, below 150~nK, no Feshbach molecules are detected in the absorption images. This disappearance coincides with the onset of  $^{41}$K Bose-Einstein condensation, and is consistent with enhanced molecular loss due to the high atomic density.

The dependence of the molecule number on RF pulse duration and mixture ratio further reflects a competition between association and inelastic loss due to collisions with residual atoms. To facilitate a quantitative analysis of these kinetics, we employ square RF pulses of constant amplitude rather than the Blackman pulses used for spectroscopy, ensuring a time-independent association rate that simplifies the modeling. The resulting molecular yields are shown in Fig.~\ref{Optimization}(b) and (c), where we vary the RF pulse duration at a controlled atom number ratio of 0.4(1) and the atom number ratio at fixed pulse duration 3.15~m, respectively. To interpret these observations, we use a rate-equation model adapted from Ref.~\cite{Voges2020Na39K} and extend it to distinguish between molecule loss channels driven by collisions with $^{23}$Na and $^{41}$K. The dynamics are described by coupled equations for the molecular population $N_\mathrm{mol}$ and the atomic populations $N_\mathrm{Na}$ and $N_\mathrm{K}$:
\setlength{\parskip}{0pt}
\begin{equation}
\small
\begin{split}
\frac{dN_{\mathrm{mol}}}{dt} = &+
    k_{\mathrm{mol}}g_{\mathrm{{Na,K}}}N_{\mathrm{Na}}N_{\mathrm{K}} - \Gamma_{\mathrm{Na}}g_{\mathrm{mol,Na}}N_{\mathrm{mol}}N_{\mathrm{Na}} \\
&-\Gamma_{\mathrm{K}}g_{\mathrm{mol,K}}N_{\mathrm{mol}}N_{\mathrm{K}}\\
    \frac{dN_{\mathrm{Na}}}{dt} = &-k_{\mathrm{mol}}g_{\mathrm{{Na,K}}}N_{\mathrm{Na}}N_{\mathrm{K}} -\Gamma_{\mathrm{Na}}g_{\mathrm{mol,Na}}N_{\mathrm{mol}}N_{\mathrm{Na}}\\
    \frac{dN_{\mathrm{K}}}{dt}  = &-k_{\mathrm{mol}}g_{\mathrm{{Na,K}}}N_{\mathrm{Na}}N_{\mathrm{K}} -\Gamma_{\mathrm{K}}g_{\mathrm{mol,K}}N_{\mathrm{mol}}N_{\mathrm{K}}
\end{split}
\label{couple}
\end{equation}

Here, $k_{\mathrm{mol}}$ is the molecule association coefficient, which depends on the RF pulse parameters, while $\Gamma_\mathrm{Na}$ and $\Gamma_\mathrm{K}$ are two-body inelastic loss coefficients for Na-NaK and K-NaK collisions, respectively. The geometric overlap factors  $g_{i,j}=\int n_i(\vec{r}) n_j(\vec{r}) d^3r/N_i N_j$ account for spatial mode matching, including the differential gravitational sag between Na-K, and are evaluated assuming Gaussian density profiles determined by the known trap parameters. 
We fit the solutions of Eq.~(\ref{couple}) simultaneously to the measured molecule numbers as functions of pulse duration [Fig.~\ref{Optimization}(b)] and atom number ratio [Fig.~\ref{Optimization}(c)]. The model parameters are optimized by minimizing the combined least-squares error, yielding $k_{\mathrm{mol}} = 1.3(2)\times10^{-10}$~cm$^3$/s, $\Gamma_\mathrm{Na} = 3.3(7)\times10^{-9}$~cm$^3$/s, and $\Gamma_\mathrm{K} = 7.4(43)\times10^{-11}$~cm$^3$/s. The model reproduces the observed data well and shows that the inelastic loss coefficient for Na-NaK collisions is about 45 times larger than that for K-NaK collisions, consistent with the resonant character of the collisions involving $^{23}$Na near the Feshbach resonance. The maximum molecule number, $N_\textrm{mol} = 1.1(1) \times 10^4$, occurs at an optimal atom number ratio of $N_{\text{Na}}$/$N_{\text{K}}$~$\approx$~0.27. The corresponding association efficiency is about 10$\%$, normalized to the minority $^{23}$Na atoms. For larger ratios ($N_{\text{Na}}$/$N_{\text{K}}$~$\gtrapprox 1.5$), the data begin to deviate from the model prediction. This discrepancy suggests the onset of higher-order loss processes not included in our two-body description, such as three-body recombination involving two sodium atoms that become increasingly significant at high $^{23}$Na densities.

 \begin{figure}[t]
    \includegraphics[width=0.48\textwidth]{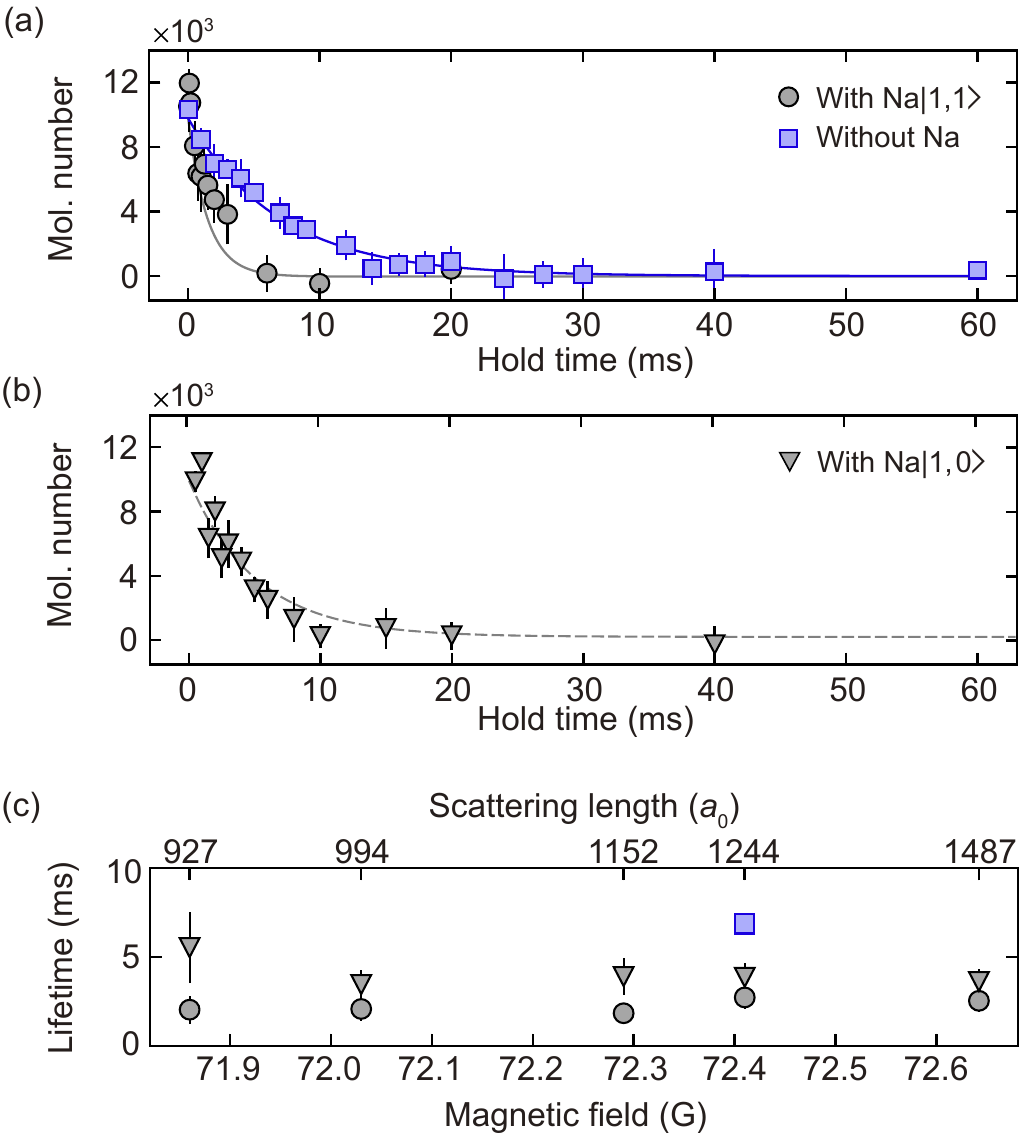}
    \caption{Lifetime measurement of $^{23}$Na$^{41}$K Feshbach molecules. (a) Gray circles show the number of molecules as a function of hold time  with $^{23}$Na prepared in the $\vert1,1\rangle$ state. Blue squares show the number of molecules with $^{23}$Na removed. The solid lines are simulations obtained from the rate-equation model using the extracted loss coefficients.
 (b) Molecular lifetime with $^{23}$Na atoms transferred to the $\vert1,0\rangle$ state. Error bars in (a) and (b) represent the standard error of the mean. (c) Measured molecular lifetimes as a function of magnetic field. The error bars of the lifetimes represent 95$\%$ confidence intervals.}
    \label{lifetime}
\end{figure}

Understanding the collisional stability of $^{23}$Na$^{41}$K Feshbach molecules is essential for efficient transfer to the rovibrational ground state. We therefore measure their lifetime in the presence and absence of background $^{23}$Na atoms, which dominate the inelastic loss. With $^{23}$Na present, the 1/$e$ lifetime of the molecular gas is 2.1(8)~ms [Fig.~\ref{lifetime}(a)]. To selectively remove the residual $^{23}$Na atoms without significantly affecting the Feshbach molecules, we transfer the sodium atoms to the $\vert1,0\rangle$ state using a 120~$\mu$s RF LZ sweep, followed by a 30~$\mu$s resonant optical clean-out pulse. After this purification, the 1/$e$ lifetime extends to 6.9(10)~ms. 

The solid curves in Fig.~\ref{lifetime}(a) are parameter-free simulations of the molecular lifetimes rather than fits. They are obtained by numerically integrating the coupled rate equations using the association and loss coefficients $k_{\mathrm{mol}}$, $\Gamma_{\mathrm{Na}}$, and $\Gamma_{\mathrm{K}}$ independently determined from the association efficiency analysis. The simulations incorporate both molecule formation and loss during the 3.15~ms association pulse, followed by purely inelastic decay after the RF field is switched off. The excellent agreement between the measured data and these simulations quantitatively validates the coupled rate-equation model and confirms that the extracted rate coefficients are predictive across different experimental conditions. 

We also measure the molecular lifetime when background $^{23}$Na is prepared in the spectator $\vert 1,0 \rangle$ state [Fig.~\ref{lifetime}(b)], obtaining a 1/$e$ lifetime of 3.8(2)~ms. Fitting the model to these data yields a Na-NaK loss coefficient of $3.5(12)\times10^{-10}$~cm$^3$/s, roughly an order of magnitude smaller than in the resonant $\vert 1,1 \rangle$ state. We further observe that these lifetimes remain nearly constant across the experimentally accessed magnetic field range [Fig.~\ref{lifetime}(c)]. This weak dependence on the scattering length reflects the broad, open-channel dominated nature of the resonance. 

Notably, the observed 6.9(10)~ms lifetime is more than an order of magnitude longer than that reported for bosonic $^{23}$Na$^{39}$K Feshbach molecules~\cite{Voges2020Na39K}. We expect this lifetime to be further extended by removing the remaining $^{41}$K atoms~\cite{Lam2022NaCs}. This enhanced stability, combined with the fact that typical STIRAP transfer times are on the order of 20--100~$\mu$s, provides a robust platform for the production of $^{23}$Na$^{41}$K ground state molecules. 

In conclusion, we have established a robust protocol for creating ultracold bosonic $^{23}$Na$^{41}$K Feshbach molecules via RF association. By performing binding energy measurements, we have pinned down the interspecies scattering properties, identifying a broad low-field resonance centered at $B_0$ = 73.6(1)~G with a width of $\Delta$ = 5.1(2)~G. This large width is particularly advantageous, offering a stable handle for tuning interactions to further optimize sympathetic cooling. In the present implementation that relies on background scattering, our experiment yields samples of $1.1(1)\times10^4$ Feshbach molecules with an estimated phase-space density of 0.03(1), comparable to starting conditions of molecular species that have been successfully condensed after ground state transfer~\cite{Bigagli2024NaCsBEC, Shi2025NaRb}.

Our quantitative analysis of association and loss dynamics establishes predictive control of molecule formation and decay. Using collision parameters derived directly from the association spectroscopy, our extended rate-equation model quantitatively reproduces the observed molecular lifetimes without additional fitting. Importantly, this analysis reveals that the present molecular yield is primarily creation rate limited. Since the association coefficient $k_\mathrm{mol}$ scales with the square of the effective Rabi coupling in the perturbative regime, increasing the RF power provides a direct route to larger molecule numbers. Furthermore, improving the spatial density overlap via a tailored potential could further enhance the association efficiency.

Together, these results provide an excellent starting point for STIRAP transfer to the rovibrational ground state and subsequent evaporative cooling. A large initial number and high phase-space density are key prerequisites for efficient evaporation, and recent demonstrations of dipolar shielding and evaporation outline a realistic pathway for extending the present work toward quantum degenerate gases of bosonic $^{23}$Na$^{41}$K, opening access to strongly dipolar many-body regimes. Combined with the creation of fermionic $^{23}$Na$^{40}$K molecules within the same experimental apparatus~\cite{Chang2025Machine}, this work establishes NaK as a versatile platform for complementary studies of bosonic and fermionic dipolar physics.\\

\section*{Acknowledgements }
This work was supported by the National Research Foundation of Korea (Grants No. RS-2025-00559423, RS-2025-02317602, RS-2023-NR119931), the Samsung Science and Technology Foundation (Project No. SSTF-BA2001-06), and the Ministry of Science and ICT, Korea, under
the ITRC support program (Grant No. RS-2022-00164799) supervised by the IITP. Y. Lim acknowledges the support from the NRF BK21 FOUR Postdoctoral Fellowship and S. Lee from the Hyundai Motor Chung Mong-Koo Foundation.

\end{document}